%% file: main.tex
\documentclass[sigconf]{acmart}
\usepackage{booktabs} 
\usepackage{graphicx}
\usepackage{algorithm}
\usepackage{algpseudocode}
\usepackage{bm}
\usepackage{balance}

\def\BibTeX{{\rm B\kern-.05em{\sc i\kern-.025em b}\kern-.08emT\kern-.1667em\lower.7ex\hbox{E}\kern-.125emX}}

\newtheorem{theorem}{Theorem}

\newtheorem{assumption}{Assumption}

\begin{document}
\title{A Unified Framework for Marketing Budget Allocation}

\author{Kui Zhao, Junhao Hua, Ling Yan, Qi Zhang, Huan Xu, Cheng Yang}
\affiliation{
  \institution{Machine Intelligence Technology Lab, Alibaba Group}
  \city{Hangzhou} 
  \state{Zhejiang} 
  \country{China}
  \postcode{311121}
}
\email{{zhaokui.zk, junhao.hjh, yanling.yl, qiyue.zq, huan.xu, charis.yangc}@alibaba-inc.com}
\renewcommand{\shortauthors}{Kui Zhao, et al.}

\begin{abstract}
While marketing budget allocation has been studied for decades in traditional business, nowadays online business brings much more challenges due to the dynamic environment and complex decision-making process. In this paper, we present a novel unified framework for marketing budget allocation. By leveraging abundant data, the proposed data-driven approach can help us to overcome the challenges and make more informed decisions. In our approach, a semi-black-box model is built to forecast the dynamic market response and an efficient optimization method is proposed to solve the complex allocation task. First, the response in each market-segment is forecasted by exploring historical data through a semi-black-box model, where the capability of logit demand curve is enhanced by neural networks. The response model reveals relationship between sales and marketing cost. Based on the learned model,  budget allocation is then formulated as an optimization problem, and we design efficient algorithms to solve it in both continuous and discrete settings. Several kinds of business constraints are supported in one unified optimization paradigm, including cost upper bound, profit lower bound, or ROI lower bound. The proposed framework is easy to implement and readily to handle large-scale problems. It has been successfully applied to many scenarios in Alibaba Group. The results of both offline experiments and online A/B testing demonstrate its effectiveness. 
\end{abstract}

%
%



\copyrightyear{2019} 
\acmYear{2019} 
\setcopyright{acmcopyright}
\acmConference[KDD '19]{The 25th ACM SIGKDD Conference on Knowledge Discovery and Data Mining}{August 4--8, 2019}{Anchorage, AK, USA}
\acmBooktitle{The 25th ACM SIGKDD Conference on Knowledge Discovery and Data Mining (KDD '19), August 4--8, 2019, Anchorage, AK, USA}
\acmPrice{15.00}
\acmDOI{10.1145/3292500.3330700}
\acmISBN{978-1-4503-6201-6/19/08}
\fancyhead{}

\maketitle

\input{introduction}

\input{market_response}
\input{budget_allocation}
\input{experiment}
\input{related_work}

\input{conclusion}

\begin{acks}
The authors would like to thank Chao Wang, Tuo Shi, Jun Lang, Yu Dong, Tianxiang Xu, Jun Tan, Zulong Chen, Ling Chen 
for discussions and supports. 
\end{acks}

\bibliographystyle{ACM-Reference-Format}
\bibliography{bibliography.bib}

\appendix
\input{appendix}

\end{document}

%% file: introduction.tex
\section{Introduction}
\begin{figure}[tbp!]
\centering
\includegraphics[width=0.81\linewidth]{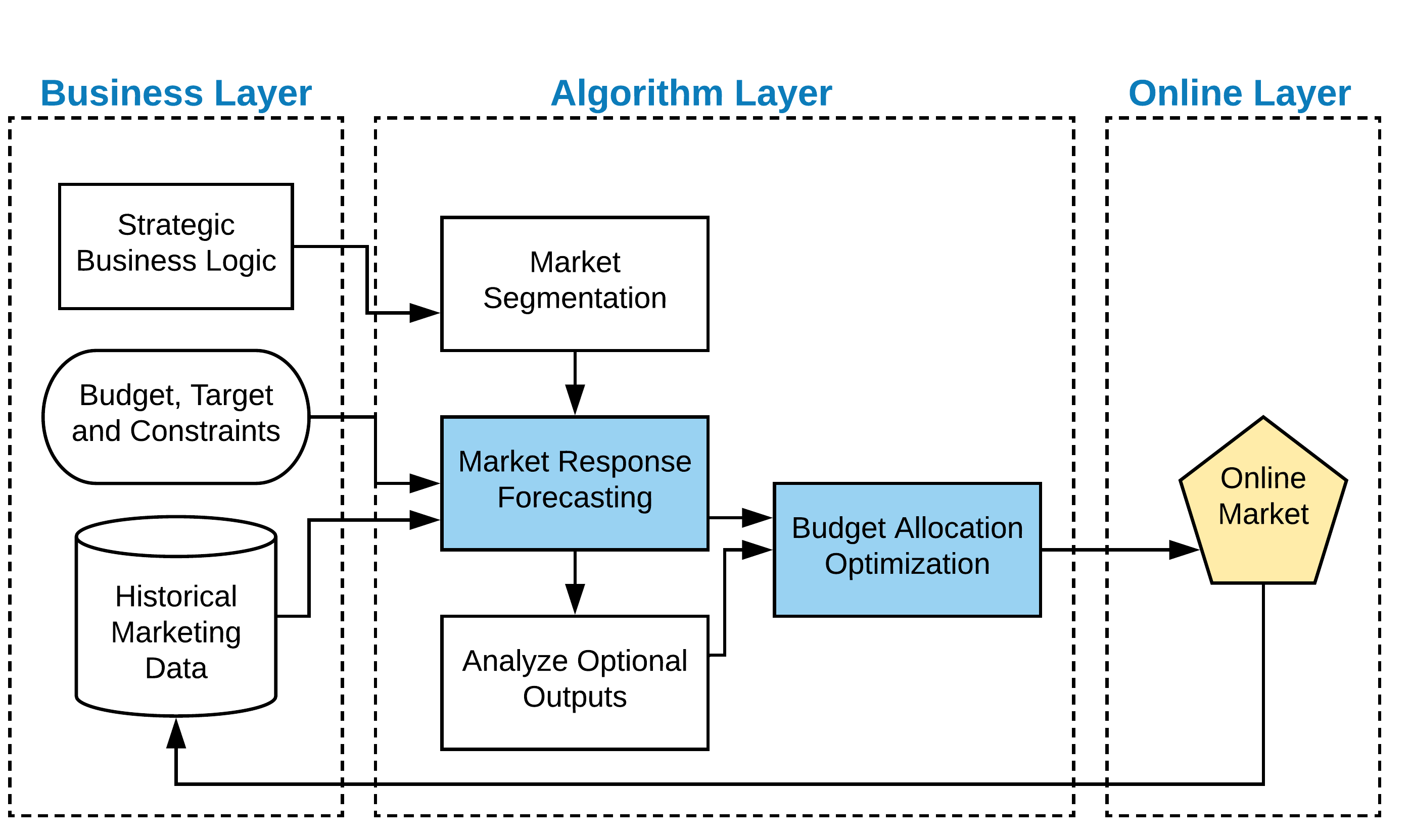} 
\caption{Our framework for marketing budget allocation.} 
\label{framework} 
\end{figure}

When and where should you spend the money? 
The answer to this question is the key to budget allocation in marketing 
and this topic has been studied for decades in traditional business. 
The environment and decision-making process in online business are much more dynamic and complex 
than traditional ones. And the marketing budget allocation is sophisticated and needs to be readjusted quite often, 
e.g. weekly or even daily, much more often than a few times per year like traditional marketing does. 
That brings great challenges, especially for companies 
with large and diversified markets \cite{laudon2016commerce, strauss2016marketing}. 
Fortunately, in online business, commercial activities can be tracked and depicted 
by the collected data from multiple sources. 
By leveraging these data, data-driven methods can help us to overcome above challenges. 
The intelligent techniques allow us to gain deeper insights into the market and make more informed decisions. 
Moreover, when the market is monitored and analyzed automatically, 
new patterns can be detected more quickly, and the budget allocation is thus readjusted timely \cite{akter2016big}. 
In this paper, we present a novel unified framework for marketing budget allocation, 
which contains two sequential steps: learning market response models from historical data, 
and optimizing budget allocation based on learned models. 
The whole framework is illustrated in \figurename { \ref{framework}}.  

In marketing, the whole market is divided into many segments, 
according to different commodities, customers, and consumption time, etc. 
In order to resolve how to allocate the budget into each market-segment, 
we need to forecast sales response of each segment if the budget is spent on it.  
Although black-box forecasting methods, such as neural networks, have been widely used in 
many applications \cite{tong2017simpler, liao2018deep, taieb2017regularization}, 
there are several gaps between forecasting and decision making \cite{athey2017beyond}. 
One of the biggest challenges is that it is too difficult 
to translate black-box forecasting into allocation decisions.
In contrast, the logit demand curve has an explicit expression 
and it is very popular in economics \cite{phillips2005pricing, qu2013learning, van2018dynamic}. 
However, fitting logit demand curves in each market-segment separately
has no capability of sharing information among different segments. 
That may cause serious data sparseness problem. 
Therefore, we propose a semi-black-box model to extend 
the capability of logit demand curves via neural networks. 
The input of forecasting is firstly split into two different parts: 
independent variables and contextual variables. 
The independent variable we have here is the budget allocated to each segment. 
The contextual variables are intrinsic attributes of market, 
such as brands of commodities, living cities of customers, consumption time and so on.
It will be shown that the elasticity at market average cost (or market cost)
of logit response model is determined by the bias parameter. 
By taking contextual variables as input, we use the same neural network 
to calculate bias parameters for all segments. 
Based on that, fitting logit demand curves is extended to  
a process of learning the elasticity at market cost 
together with a specific sensitivity coefficient for each segment, 
where the relationship between sales and cost is depicted in an explicit way.  

As long as response models are acquired, budget allocation can be 
formulated as an optimization problem. 
Let $N$ denote the number of market-segments, 
budget allocation needs to optimize over $N$ variables 
to maximize sales under some budget constraint. 
Due to the non-convexity of logit demand curves, 
the optimization problem is non-convex.
Inspired by \cite{dong2009dynamic, li2011pricing}, 
we reformulate the problem into an equivalent convex optimization problem. 
By introducing a dual variable, the Lagrange multiplier method is applied. 
According to KKT conditions, 
the original problem is reduced to a root finding problem 
with respect to dual variable, 
and it can be solved by bisection on one dual variable, 
instead of searching $N$ primal variables. 
The algorithm converges in a few iterations and the complexity of each iteration is $O(N)$, 
thus it is readily to handle large-scale problems. 
Several kinds of business constraints, including cost upper bound, profit lower bound, 
or ROI lower bound, are supported in one unified paradigm with minor adaptations. 
What's more, the optimization paradigm can be easily extended to discrete settings
with a little additional computation, 
where budget variables can only be chosen from a set of discrete values. 
One major cause of discrete decision variables is that 99-ending display prices, e.g. \$99, \$199 or \$299, 
are usually preferred after discount or premium \cite{schindler1996increased} in many situations. 
We will show that the discrete setting can be solved by 
constructing a Multiple Choice Knapsack Problem (MCKP) \cite{kellerer2004multiple} 
from the solution of relaxed continuous problem. 
Dyer et al. \cite{dyer1984n} and Zemel et al. \cite{zemel1984n} independently 
developed approximation algorithms for MCKP, running in $O(N)$ time, 
and the algorithm can be further boosted by pruning \cite{pisinger1995minimal}.

The proposed framework is easy to implement and 
has been successfully applied to many 
scenarios in 
Alibaba Group\footnote{
Taopiaopiao is a platform for selling movie tickets.
Koubei and Ele.me are platforms for local services.
CBU (also called 1688) is a business-to-business (B2B) platform. 
Fliggy is a platform for selling travel tickets.
Tmall is a business-to-consumer (B2C) platform, 
and Lazada is the Southeast Asian version. 
Credit Pay (or Huabei) provides virtual credit card service, 
and Cash Now (or Jiebei) provides consumer loan service. 
}: 
\begin{itemize}
\item {\it Daily Marketing}: Taopiaopiao, Koubei, Ele.me, CBU, Fliggy.
\item {\it Shopping Festival}: 11.11 in Tmall and 12.12 in Lazada.
\item {\it Interest Discount}: Credit Pay and Cash Now in Ant Financial. 
\end{itemize}
To demonstrate the effectiveness of our framework, 
we present results of both offline experiments and online A/B testing.
The offline part includes experiments on public datasets and synthetic simulations. 
For the online part, we would like to take the daily marketing in Taopiaopiao as a typical example, 
on which an A/B testing is conducted for five weeks and the improvements are remarkable. 
In addition, it is worth mentioning that there are billions of decision variables 
in several scenarios, such as Figgy, Credit Pay, Cash Now, etc., 
and our approach can handle them very well. 

The rest of our paper is organized as follows. 
We describe how to learn the market response in Section 2, 
and how to allocate budget in Section 3. 
We present the experimental setups and results in Section 4.
We briefly review related work in Section 5. 
The conclusions and future plans are given in Section 6. 
Finally, details related to reproducibility are given in Supplement Section.

%% file: market_response.tex
\section{Market Response}
In online business, the commercial activities can be tracked 
and depicted by collected data from multiple sources. 
Based on that data, the whole market can be divided into more granular segments 
than traditional ones. Each segment may consist of  
very specific commodities and customers. For instance, 
black wool sweater for children is a market-segment. 
The market can be further segmented according to consumption time, 
e.g. movie tickets on Saturday evening at a specific cinema is 
another example of market-segment. 

\subsection{Logit Response Model}
The fundamental building block of marketing budget allocation is to 
forecast sales response of each segment if the budget is spend on it. 
As black-box forecasting methods, neural networks are very powerful 
and have wide applications \cite{tong2017simpler, liao2018deep, taieb2017regularization}. 
However, there are many challenges to translate black-box forecasting into 
allocation decisions \cite{athey2017beyond}. 
In contrast, the demand curves in economics usually have explicit expressions. 
There are many kinds of demand curves, such as linear, log-linear, 
constant-elasticity, logit and so on \cite{phillips2005pricing, talluri2006theory}. 
In this work, we focus on logit demand curve, 
which is the most popular response 
model \cite{phillips2005pricing, qu2013learning, van2018dynamic}.  
\begin{figure}[tbp!]
\centering
\includegraphics[width=0.71\linewidth]{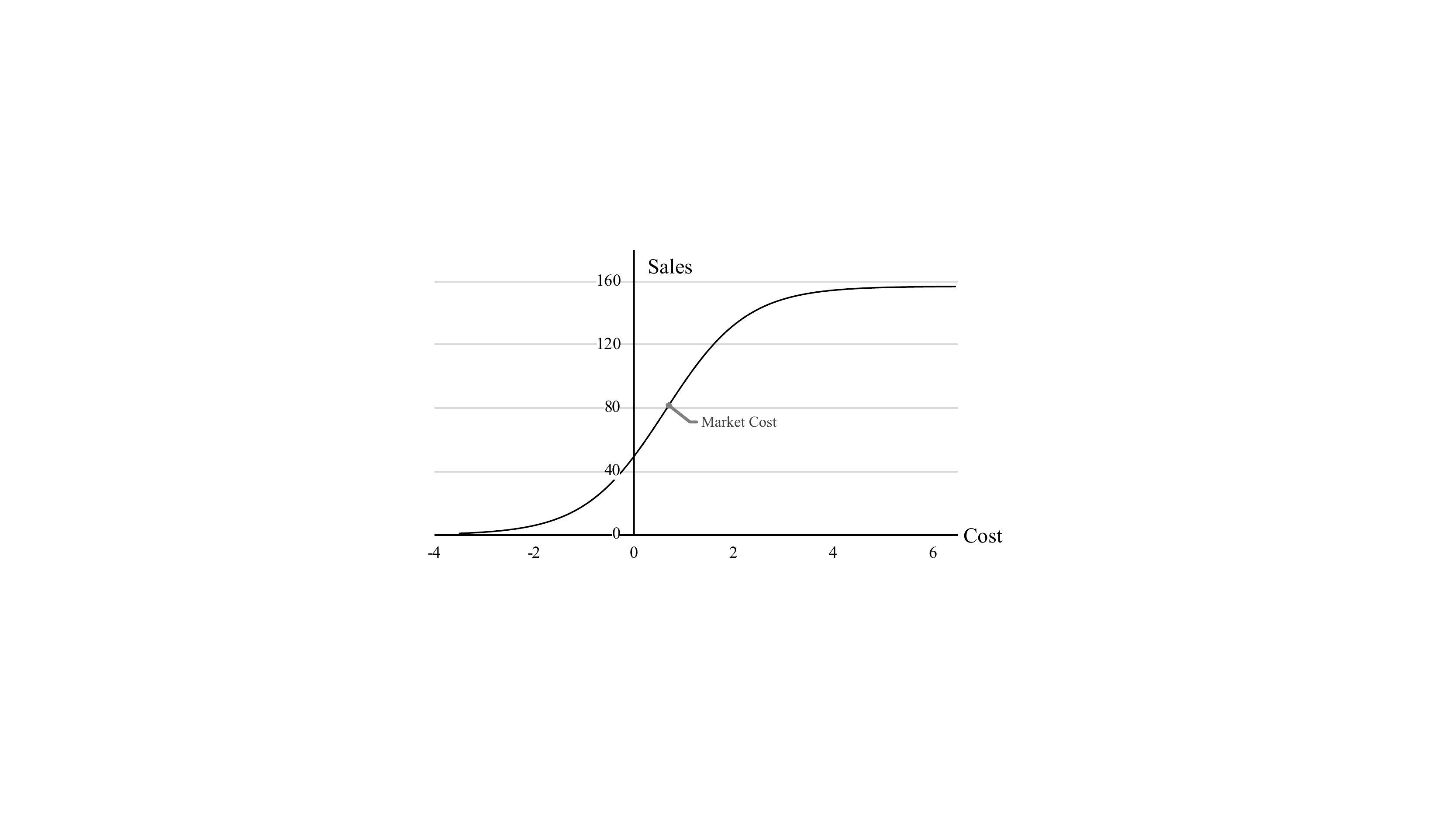} 
\caption{The logit response model.} 
\label{fig:logit} 
\end{figure}

The logit model for $i$-th segment is defined as follow 
and one sample is illustrated in \figurename { \ref{fig:logit}}: 
\begin{equation}
\label{eq:logit}
d_i(c) := \frac{D_i}{1+\exp{\{-(a_i + b_i c)\}}}, i=1,\dots, N,
\end{equation}
where $c$ is the unit cost on marketing, and $D_i > 0$ indicates the overall size of $i$-th market-segment. 
$c$ can be either greater than or less than $0$. Positive cost on marketing means 
price discount and negative cost means price premium. 
Since the size of each market-segment is quite stable, 
$D_i$ can be calculated by simple statistics in advance. 
So the only parameters need to be learned are $a_i, b_i$. 
Specifically, $b_i$ indicates sensitivity and in normal situations we have $b_i > 0$. 
That is to say greater cost on marketing brings higher sales. 
The response curve is steepest at point $\hat{c}_i = - a_i / b_i$ 
(as indicated in \figurename{ \ref{fig:logit}}), and we call this point as {\it market cost}, 
which represents the average cost on marketing (including all competitors) in the whole $i$-th market-segment. 
The sales is very sensitive to cost when the cost is close to $\hat{c}_i$. 
In other words, a substantial shifts in sales could be expected even if 
the change in cost is very small. 
The sales becomes less sensitive to cost when cost 
goes away from market cost, which is called as 
{\it diminishing marginal effect}. 

The most common measure of sensitivity is elasticity. 
Inspired by {\it price elasticity} \cite{phillips2005pricing}, we define {\it cost elasticity} as the ratio of 
percentage change in sales to percentage change in cost: 
\begin{equation}
e_i(c_{1}, c_{2}):=\frac{[d_i(c_2)-d_i(c_1)]c_1}{[c_2 - c_1]d_i(c_1)}.
\label{elasticity1}
\end{equation}
The elasticity at $c$ can be derived by taking the limit of Eq. (\ref{elasticity1})
as $c_2$ approaches $c_1$: 
\begin{equation}
e_i(c)=\nabla_c d_i(c)\frac{c}{d_i(c)}.
\label{elasticity2}
\end{equation}

\begin{theorem} \label{theor_elasticity}
The elasticity at market cost $\hat{c}_i = - a_i / b_i$ for logit response model 
is equal to $-a_i/2$. 
\footnote{The proofs of all theorems can be found in the supplement.}
\end{theorem}

\subsection{Semi-black-box Model}
The capability of fully utilizing data for logit model is limited. More specifically, 
when we learn $a_i$ and  $b_i$ for each market-segment $i$ separately, 
there is no capability of sharing information across different segments. 
For market-segments with historical budget allocation of few different values, 
that may cause serious data sparseness problem. 
Actually, the input variables of forecasting model can be split into two different parts: 
independent variables and contextual variables. 
The independent variable we have here is $c$ or the budget allocated to each segment. 
The contextual variables are intrinsic attributes of market and we denote it as $\bm{x}_i$, 
which may include brands of commodities, categories of commodities, 
living cities of customers, and consumption time, etc. 
While contextual variables vary across different segments, 
they are not changing with the independent variable in each segment.  
Traditional neural networks take both $\bm{x}_i$ and $c$ as input, 
and link them with sales in a black-box way. 
Contrarily, the logit model only takes $c$ as input, 
and link it with sales in an explicit way. 
We now propose a semi-black-box model, 
which can utilize both $\bm{x}_i$ and $c$ as input, 
as well as keep the relationship between $c$ and sales in an explicit.  

\begin{assumption} \label{assumption_elasticity}
The elasticity at market cost for logit response model is determined by contextual variables.  
\end{assumption}
It is intuitive and reasonable to introduce the above assumption. 
For instance, T-shirts are less elastic than sweaters in summer, 
and movie tickets on the weekend is less elastic than that on the weekday. 
Instead of directly forecasting the final sales, we use a neural network $e(\bm{x_i})$ 
to estimate the elasticity at market cost, i.e. $-a_i/2$, for each market-segment $i (i=1,\dots, N)$. 
Based on $e(\bm{x_i})$, the elasticity information can be shared across different segments. 
Then the traditional logit response model is extended to a semi-black-box model:
\begin{equation}
\label{eq:logit_new}
d_i(c) = \frac{D_i}{1+\exp{\{2 e(\bm{x}_i) - b_i c\}}}, 
\end{equation}
which integrates the capability of neural networks 
and explicitness of logit demand curves. 

The parameters in neural network $e(\bm{x}_i)$ as well as $b_i (i=1,\dots, N)$ 
need to be learned. We first define {\it market share} $q_i(c)$ as follow:
\begin{equation}
q_i(c):=\frac{d_i(c)}{D_i}=\frac{1}{1+\exp{\{-(a_i + b_i c)\}}},
\end{equation}
where $a_i = -2e(\bm{x}_i)$. As $D_i$ can be calculated by simple statistics in advance, 
it is easy to construct training set $\{([\bm{x}_i, c_1], \hat{q}_{i1}), \cdots, \\ ([\bm{x}_i, c_{M_i}], \hat{q}_{iM_i})\}$ 
for each market-segment $i(i=1,\dots, N)$ from historical observations. 
Then parameters are learned by minimizing the following 
negative log-likelihood through gradient decent: 
\begin{equation}
\mathcal{J}=-\frac{1}{\sum_{i=1}^{N}M_i}\sum\limits_{i=1}^{N}\sum\limits_{j=1}^{M_i}[\hat{q}_{ij} \log q_{ij} + (1-\hat{q}_{ij}) \log (1-q_{ij})].
\label{eq:ll}
\end{equation}

%% file: budget_allocation.tex
\section{Budget Allocation}
As long as response models are acquired, 
budget allocation can be formulated as an optimization problem. 

\subsection{Problem Formulation}
Suppose there are $N$ market-segments, budget allocation needs to decide unit cost $c_i$ for each 
segment $i (i=1,\dots, N)$ under some budget constraint. Let $\bm{c}:=[c_1, \dots, c_N]$, 
our objective is to find an optimal $\bm{c}$ maximizing total sales under budget constraint. Formally,  
\begin{equation} \label{obj_func_1}
\begin{aligned}
\min_{\bm{c} } & -\sum_{i =1}^N d_i(c_i), \\
\text{s.t.} \  & \sum_{i=1}^N d_i(c_i) c_i \le B,
\end{aligned}
\end{equation}
where $B$ is the budget bound and it can be either greater than or less than 0. 
Positive $B$ means cost upper bound, and negative $B$ means profit lower bound. 

Because $\sum_{i=1}^N d_i(c_i) c_i=\sum_{i=1}^N D_i q_i(c_i) c_i$ is non-convex with respect to $c_i$, 
directly solving the above problem is difficult. 
Inspired by \cite{dong2009dynamic, li2011pricing}, we can reformulate the problem 
into an equivalent convex optimization problem. 
Since $q_i(c_i)$ obviously is a strictly increasing function when $b_i > 0$, 
we can get its inverse function, which maps 
market share $q_i$ to marketing cost $c_i$, as follow: 
\begin{equation} \label{cost_eq}
\begin{aligned}
c_i(q_i) &= -\frac{a_i}{b_i} - \frac{1}{b_i} [ \ln(1-q_i) - \ln q_i],\\
q_i & \in (0, 1), i=1,\dots,N.
\end{aligned}
\end{equation}
Then we define $\bm{q}:=[q_1,\dots, q_N]$ and constraint function $g(\bm{q})$ as: 
\begin{equation}\label{con_fun}
g(\bm{q}) := \sum\limits_{i=1}^N D_i q_i c_i(q_i) - B.
\end{equation}

\begin{theorem} \label{theor_q_convex}
$g(\bm{q})$ is strongly convex with respect to $\bm{q}$. 
\end{theorem}

Therefore, we can reformulate the original problem as follow: 
\begin{equation} \label{transform_obj}
\begin{aligned}
\min_{\bm{q}} & -\sum\limits_{i =1}^N D_i q_i, \\
\text{s.t.} \  & g(\bm{q}) \le 0, \\
\end{aligned}
\end{equation}
which is a convex optimization problem with respect to $\bm{q}$. 

\subsection{Algorithm}
\label{sec:alg}
By introducing a dual variable $\lambda$, the Lagrange multiplier method is applied
to solve the convex optimization problem in Eq. (\ref{transform_obj}): 
\begin{equation}
\begin{aligned}
\mathcal{L}(\bm{q}, \lambda)  
& =  -\sum_{i =1}^N D_i q_i + \lambda ( \sum_{i=1}^N D_i q_i  c_i(q_i) -B ) \\
& = \sum_{i =1}^N  (\lambda D_i q_i c_i(q_i) - D_i q_i) - \lambda  B.
\end{aligned}
\end{equation}

\subsubsection{KKT conditions}
Both objective function and constraint functions are differentiable and convex, 
and the KKT conditions are: 
\begin{subequations} \label{kkt_condition}
	\begin{align}
	g(\bm{q}^*)  & \le 0, \\
	\lambda^* & \ge 0,  \\
	\lambda^* g(\bm{q}^*)  & = 0, \label{kkt_dual} \\
	\nabla_{q_i}(\lambda^* D_i q_i^* c_i(q_i^*) -  D_i q_i^*) & = 0,  i=1,..., N,  \label{kkt_primal}
	\end{align}
\end{subequations}
where $\bm{q}^*, \lambda^*$ are the optimal solutions of primal and dual problems respectively, 
with zero duality gap \cite{boyd2004convex}.

While $g(\bm{q})$ is strongly convex, the minimum value of $g(\bm{q})$ exists, 
and let $g(\tilde{\bm{q}})$ denote it. 
To obtain the minimum value, we take the partial derivative of $g(\bm{q})$ 
with respect to each $q_i$ and set them equal to $0$. 
Then, we can get:
\begin{equation}
\ln \frac{\tilde{q}_i}{1-\tilde{q}_i}+ \frac{\tilde{q}_i}{1-\tilde{q}_i} = a_i -1, i=1,\dots,N.
\end{equation}
Using Lambert $W$ function \cite{corless1996lambertw}, 
$g(\tilde{\bm{q}})$ is {\it uniquely} attained at: 
\begin{equation}
\tilde{q}_i = \frac{W(\exp(a_i-1))}{W(\exp(a_i-1)) + 1}, i=1,\dots,N,
\end{equation}
where $W(\cdot)$ is the Lambert $W$ function. When $g(\tilde{\bm{q}})>0$, there is no solution. 
When $g(\tilde{\bm{q}})=0$, the optimal solution is imminently obtained at $\bm{q}^* = \tilde{\bm{q}}$. 
When $g(\tilde{\bm{q}})<0$, the following assumption holds: 
\begin{assumption} \label{assumption_slater}
There exists $\bm{q}$ such that $g(\bm{q}) < 0$.
\end{assumption}

When the above assumption holds, the constraint function $g(\bm{q})$ satisfies Slater's condition, 
and the KKT conditions in Eq. (\ref{kkt_condition}) provide necessary 
and sufficient conditions for optimality \cite{boyd2004convex}. 

\begin{theorem} \label{theor_kkt_equvi}
	The fourth KKT condition in Eq. (\ref{kkt_primal}) is equivalent to $\lambda^* \neq 0$ and 
	$q_i^*= \frac{W \big(\exp (a_i + \frac{b_i}{\lambda^*} - 1) \big)}{W \big(\exp (a_i + \frac{b_i}{\lambda^*} - 1) \big)+1}(i=1, \dots, N)$, 
	where $W(\cdot)$ is the Lambert $W$ function.
\end{theorem}

From $\lambda^* \neq 0$ and Eq. (\ref{kkt_dual}), it is easy to conclude that $g(\bm{q^*}) =0$. 
Then we obtain the following optimality conditions:
\begin{subequations} \label{optimal_condition}
	\begin{align}
	g(\bm{q}^*)  & = \sum_{i=1}^N D_i q_i^* c_i(q_i^*) - B = 0, \label{new_kkt_primal} \\
	q_i^* & = \frac{W \big(\exp (a_i + \frac{b_i}{\lambda^*} - 1) \big)}{W \big(\exp (a_i + \frac{b_i}{\lambda^*} - 1) \big)+1}, i=1, \dots, N,  \label{new_kkt_dual} \\
	\lambda^* & > 0.
	\end{align}
\end{subequations}
We can see that the primal variables $q_i^*(i=1,\dots,N)$ can be represented by the same dual variable $\lambda^*$. 
That is to say searching $N$ primal variables is equal to searching one dual variable $\lambda$. 
Let us define: 
\begin{equation}\label{q_lambda}
\begin{aligned}
{q}_i(\lambda) &:= \frac{W \big(\exp (a_i + \frac{b_i}{\lambda} - 1) \big)}{W \big(\exp (a_i + \frac{b_i}{\lambda} - 1) \big)+1}, \\
\lambda &\in (0, +\infty),  i=1, \dots, N,
\end{aligned}
\end{equation}
and 
\begin{equation}
\begin{aligned}
f(\lambda) &:= \sum_{i=1}^{N} D_i q_i(\lambda),\\
g(\lambda) &:= g(\bm{q}(\lambda)),\lambda \in (0, +\infty). 
\end{aligned}
\end{equation}
If and only if the root $\lambda^*$ of $g(\lambda) = 0$ is found, 
the problem is solved and the optimal solution is $\bm{q}^* = \bm{q}(\lambda^*)$.

\subsubsection{Bisection method}
The bisection method is a numerical root-finding method, 
and the precondition is to know two values with opposite signs. 
The process repeatedly bisects the interval defined by these two values and 
breaks when the interval is sufficiently small. 
\begin{theorem}\label{theor_fg_decreasing}
The functions $f(\lambda), g(\lambda)$ are strictly decreasing with respect to $\lambda$, and 
$\lim\limits_{\lambda \to 0} g(\lambda) > 0$.
\end{theorem}

\begin{figure}[!t]
	\centering
	\includegraphics[width=0.69\linewidth]{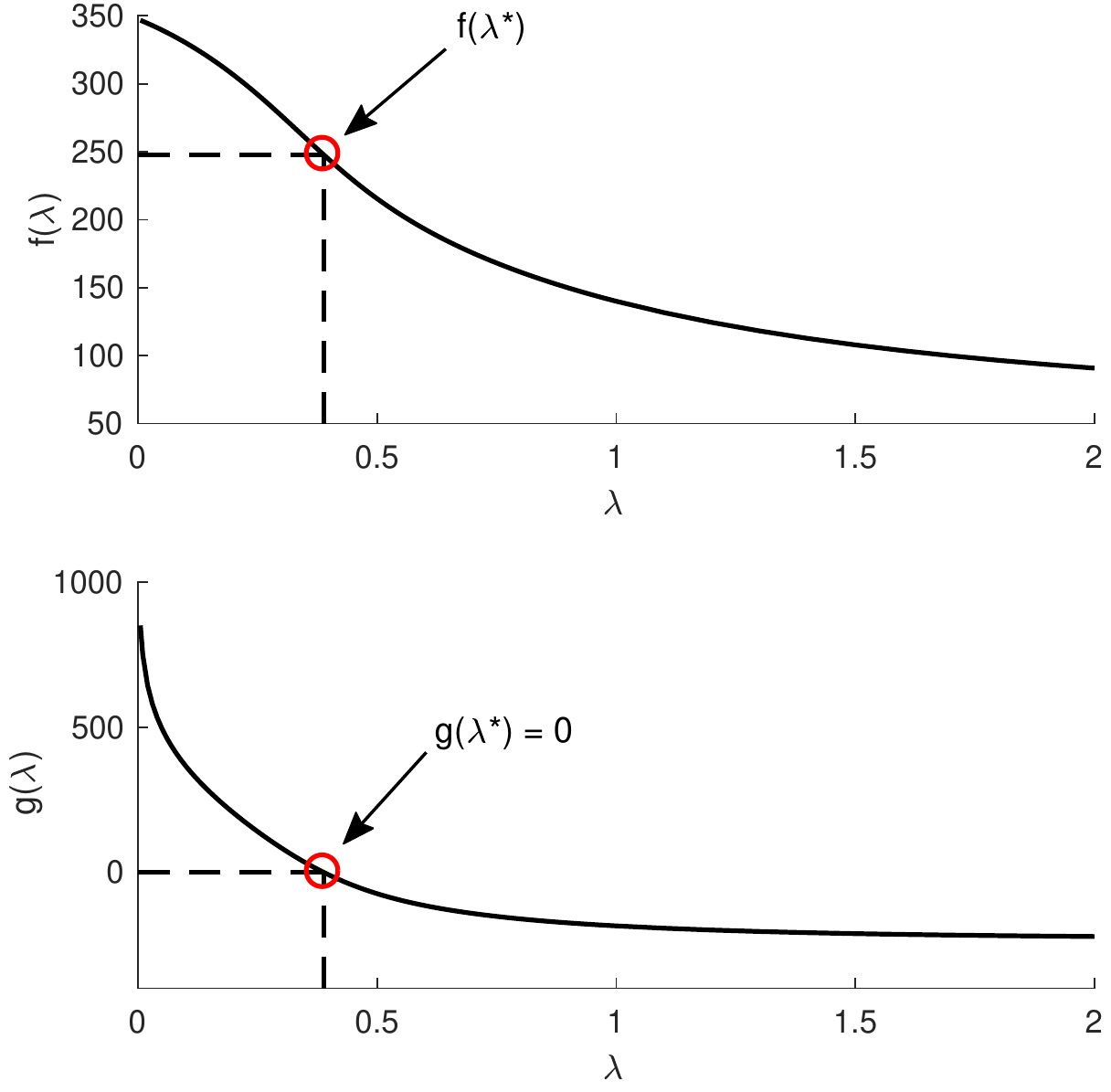}
	\caption{The plots of objective and exceeded budget with respect to dual variable $\lambda$. 
	The optimal objective value $f(\lambda^*)$ is attained when $g(\lambda^*) = 0$.}
	\label{fig:fg_lambda}
\end{figure}

When Assumption \ref{assumption_slater} holds, 
there exists $\lambda < +\infty$ such that $g(\lambda) \le 0$. 
Together with Theorem \ref{theor_fg_decreasing}, 
the root $\lambda^*$ of $g(\lambda) = 0$ 
can be found by bisection,
which is shown in Algorithm \ref{alg_bisection}, and one example is illustrated in \figurename{ \ref{fig:fg_lambda}}. 
As seen from above, we approximately set $\lambda^* = \lambda_r$ 
in order to satisfy the constraint.
In practice, the algorithm can also be early terminated 
by $f(\lambda_l) - f(\lambda_r) \le \epsilon'$, besides $\lambda_r - \lambda_l \leq \epsilon$. 
It is obvious that $f(\lambda^*) - f(\lambda_r) < \epsilon'$ 
when $f(\lambda_l) - f(\lambda_r) \le \epsilon'$. 

\begin{algorithm}[t!]
	\caption{Bisection algorithm for budget allocation} \label{alg_bisection}
	\begin{algorithmic}[1]
		\Require{ $\epsilon, B, a_i, b_i, D_i, i=1,\dots,N$.}
		\Statex   
		
		\State $\lambda_l = 0, \lambda_r = 1$.
		\While{$g(\lambda_r) > 0$}
		\State $\lambda_l = \lambda_r$, $\lambda_r = 2 \lambda_r$.
		\EndWhile
		\Repeat
		\State $\lambda_{m} = (\lambda_l +\lambda_r)/2$.
		\If{$g(\lambda_{m})>0$}
		\State $\lambda_l = \lambda_{m}$.
		\Else
		\State $\lambda_r = \lambda_{m}$.
		\EndIf
		\Until{$\lambda_r - \lambda_l \le \epsilon$}
		\State $\lambda^* = \lambda_r$, $\bm{q}^*=\bm{q}(\lambda^*)$.
		\State Compute $c_i^*=c_i(q_i^*)$ using Eq. (\ref{cost_eq}).
	\end{algorithmic}
\end{algorithm}

\subsubsection{Time complexity}
At line $2-4$ of Algorithm \ref{alg_bisection}, there are at most 
$\lceil \log_2( \max(\lambda^*, 1)) \rceil + 1$ iterations
to find $\lambda_r$ satisfying $g(\lambda_r) \leq0$.
At line $5-12$,
there are at most $ \lceil \log_2 (\frac{\max(\lambda^*, 1)}{\epsilon}) \rceil$ iterations. 
Therefore, the total number of iterations is less than 
$\lceil 2\log_2(\max(\lambda^*, 1))+ \log_2(\frac{2}{\epsilon}) \rceil$. 
If the algorithm can be further early terminated 
by $f(\lambda_l) - f(\lambda_r) \le \epsilon'$, 
there are at most $\lceil \log_2 \big(\min(\frac{\max(\lambda^*, 1)}{\epsilon}, 
\frac{\sum_{i=1}^{N}D_i}{\epsilon'})\big)\rceil$ 
iterations at line $5-12$. So the total number of iteration is less than a constant, 
i.e. $\lceil \log_2\big(\max(
\frac{1}{\epsilon}, \frac{\sum_{i=1}^{N}D_i}{\epsilon'})\big) + 1\rceil$. 
In practice, the algorithm converges in a few iterations 
and the complexity of each iteration is $O(N)$, 
thus it is readily to handle large-scale problems.

\subsection{Extension}
\subsubsection{ROI}
In many situations, it is not easy to determine the specific $B$. 
Instead, the efficiency ratio of spending budget 
is expected to be no less than a given value. 
Let $R > 0$ denote the lower bound of efficiency ratio or 
Return on Investment (ROI), the original problem in Eq. (\ref{obj_func_1}) is extended to: 
\begin{equation} \label{obj_func_roi}
\begin{aligned}
\min_{\bm{c} } & -\sum_{i =1}^N d_i(c_i), \\
\text{s.t.} \  & \sum\limits_{i=1}^N \big(R d_i(c_i) c_i -  d_i(c_i)\big) \le 0. \\
\end{aligned}
\end{equation}
Then the constraint function in Eq. (\ref{con_fun}) can be replaced by:
\begin{equation}\label{con_fun_roi}
g'(\bm{q}) := \sum\limits_{i=1}^N\big( R D_i q_i c_i(q_i) - D_i q_i\big).
\end{equation}

\begin{theorem} \label{theor_ext_convex}
$g'(\bm{q})$ is strongly convex with respect to $\bm{q}$.  
$g'(\lambda)$ is strictly decreasing with respect to $\lambda$, 
and $\lim_{\lambda \to 0} g'(\lambda) > 0$.
\end{theorem}

After a similar derivation, we can conclude that the new optimization problem 
still can be solved by Algorithm \ref{alg_bisection}, 
except $q_i(\lambda)$ in Eq. (\ref{q_lambda}) is replaced by: 
\begin{equation}
\begin{aligned}
{q}_i(\lambda) &:= \frac{W \big(\exp (a_i + \frac{b_i}{\lambda R} + 
\frac{b_i}{R} - 1) \big)}{W \big(\exp (a_i + \frac{b_i}{\lambda R} + \frac{b_i}{R} - 1) \big)+1}, \\
\lambda &\in (0, +\infty),  i=1, \dots, N,  
\end{aligned}
\end{equation}
and the input $B$ is no longer required. 

\subsubsection{Discrete settings} \label{sec_discrete_setting}
In many situations, each $c_i$ can only be chosen from a set of discrete values. 
One common reason is that 99-ending display prices
(e.g. \$99, \$199 or \$299) are preferred after discount or premium \cite{schindler1996increased}. 
Let $S_i := \{c_i^1,\dots,c_i^{|S_i|} \}$ denote the set of optional values for $c_i$, the original problem in Eq. (\ref{obj_func_1}) is then extended to a NP-hard problem: 
\begin{equation} \label{obj_func_dis}
\begin{aligned}
\min_{\bm{c} } & -\sum_{i =1}^N d_i(c_i), \\
\text{s.t.} \  & \sum_{i=1}^N d_i(c_i) c_i \le B, \\
c_i & \in S_i, i=1,\dots,N. 
\end{aligned}
\end{equation}

To solve the discrete problem, we first solve the relaxed continuous 
problem in Eq. (\ref{obj_func_1}) and let $\bm{c}^* = [c_1^*, \dots, c_N^*]$ 
denote the solution. After defining: 
\begin{equation}
\begin{aligned}
c_i^l & := \arg\min_{c_i \in S_i, c_i \le c_i^*} |c_i-c_i^*|,\\
c_i^u & := \arg\min_{c_i \in S_i, c_i > c_i^*} |c_i-c_i^*|, i=1,\dots,N.\\
\end{aligned}
\end{equation}
The discrete problem can be then solved by constructing a 
Multiple Choice Knapsack Problem (MCKP) \cite{kellerer2004multiple}, 
where $B$ is the {\it capacity} of knapsack. In the constructed MCKP, 
$i$-th class represents $i$-th market-segment, 
and each class contains two items: 
the {\it profit} and {\it weight} of the first item are $d_i(c_i^l), d_i(c_i^l)c_i^l$; 
the {\it profit} and {\it weight} of the second item are $d_i(c_i^r), d_i(c_i^r)c_i^r$. 
Since $\sum_{i=1}^N d_i(c_i^l) c_i^l \le \sum_{i=1}^N d_i(c_i^*) c_i^*  \le B$, 
there exists a solution to the constructed MCKP. 
Dyer et al. \cite{dyer1984n} and Zemel et al. \cite{dyer1984n} independently 
developed approximation algorithms (called Dyer-Zemel) for MCKP, running in $O(N)$ time, 
and the algorithm can be further boosted by pruning \cite{pisinger1995minimal}.
So the additional computation is very limited. In practice, when $|S_i|$ is small 
we suggest to directly solve the discrete setting by Dyer-Zemel algorithm
with multiple items in each class, instead of solving the relaxed continuous problem first. 
Actually, when discrete setting is directly solved by Dyer-Zemel algorithm, 
profit and weight can be generated by arbitrary methods. 
Moreover, the discrete settings under ROI constrains can be readily supported 
with few adaptations. 
 

%% file: experiment.tex
\section{Experiment}
To demonstrate the effectiveness of our approach, 
we present results of both offline experiments and online A/B testing. 

\subsection{Market Response}
\subsubsection{Datasets}
The proposed semi-black-box response model is tested on two datasets. 
\begin{itemize}
\item {\bf Breakfast.} It is a public dataset\footnote{https://www.dunnhumby.com/careers/engineering/sourcefiles} 
and contains sales and promotion records on 4 thousand market-segments 
(58 products from 77 stores) over 156 weeks. 
For each market-segment, the sales volume is accumulated weekly, 
and the market size is approximated by maximum sales in history. 
Cost on marketing (discount or premium) can be calculated from the difference 
between {\it shelf price} and {\it base price}. 
Contextual variables include store, product, manufacture, category, 
sub-category, in-store display, etc. 
\item {\bf Taopiaopiao.} It is collected from Taopiaopiao\footnote{https://www.taopiaopiao.com}, 
which is one of the largest platforms for selling movie tickets in China. 
There are 60 thousand market-segments (each day-of-week of 8 thousand cinemas) over 152 days. 
For each market-segment, the sales volume is accumulated daily, 
and the market size is provided by government agency\footnote{https://zgdypw.cn}. 
Cost on marketing can be obtained directly, 
and contextual variables include cinema, day-of-week, city, etc. 
\end{itemize}
\subsubsection{Setups}
As we can see, all contextual variables are indicator variables. 
In our model, they are firstly represented by one-hot encodings 
and then concatenated together as input layer of the neural network part. 
After that, there are 5 fully connected layers with ReLU as the activation function, 
and the dimension of each layer is 16. 
In both the baseline (logit model) and the proposed semi-black-box model, 
all trainable parameters are initialized as zero and updated through 
gradient decent with Adam rule \cite{kinga2015method}. 
The learning rate is 0.01 and all models are trained for 200 epochs. 

\subsubsection{Results} 
The results are measured in Relative Mean Absolute Error (RMAE), 
which is calculated as:  
\begin{equation}
\text{RMAE}=\frac{\sum_{i=1}^{N}|y_i-\hat{y}_i|}{\sum_{i=1}^{N}\hat{y}_i},
\end{equation}
where $N$ is the number of test samples, $\hat{y}_i$ is the true value and $y_i$ is the prediction value. 
For both Breakfast and Taopiaopiao, the whole datasets are split into 
training and test sets along business timeline. 
In order to comprehensively evaluate our method, we measure performances under different partition ratios. 
In particular, the proportion of dataset selected as training set ranges from 5\% to 95\%, 
and all results are plotted in \figurename{ \ref{fig:tpp_breakfast}. 
As can be seen, our semi-black-box model significantly outperforms the traditional logit model, 
especially when the training data is limited. The reason is that our model is able to 
share information across different segments, 
and thus data sparseness problem is greatly relieved. 

\begin{figure}[!t]
	\centering
	\includegraphics[width=0.93\linewidth]{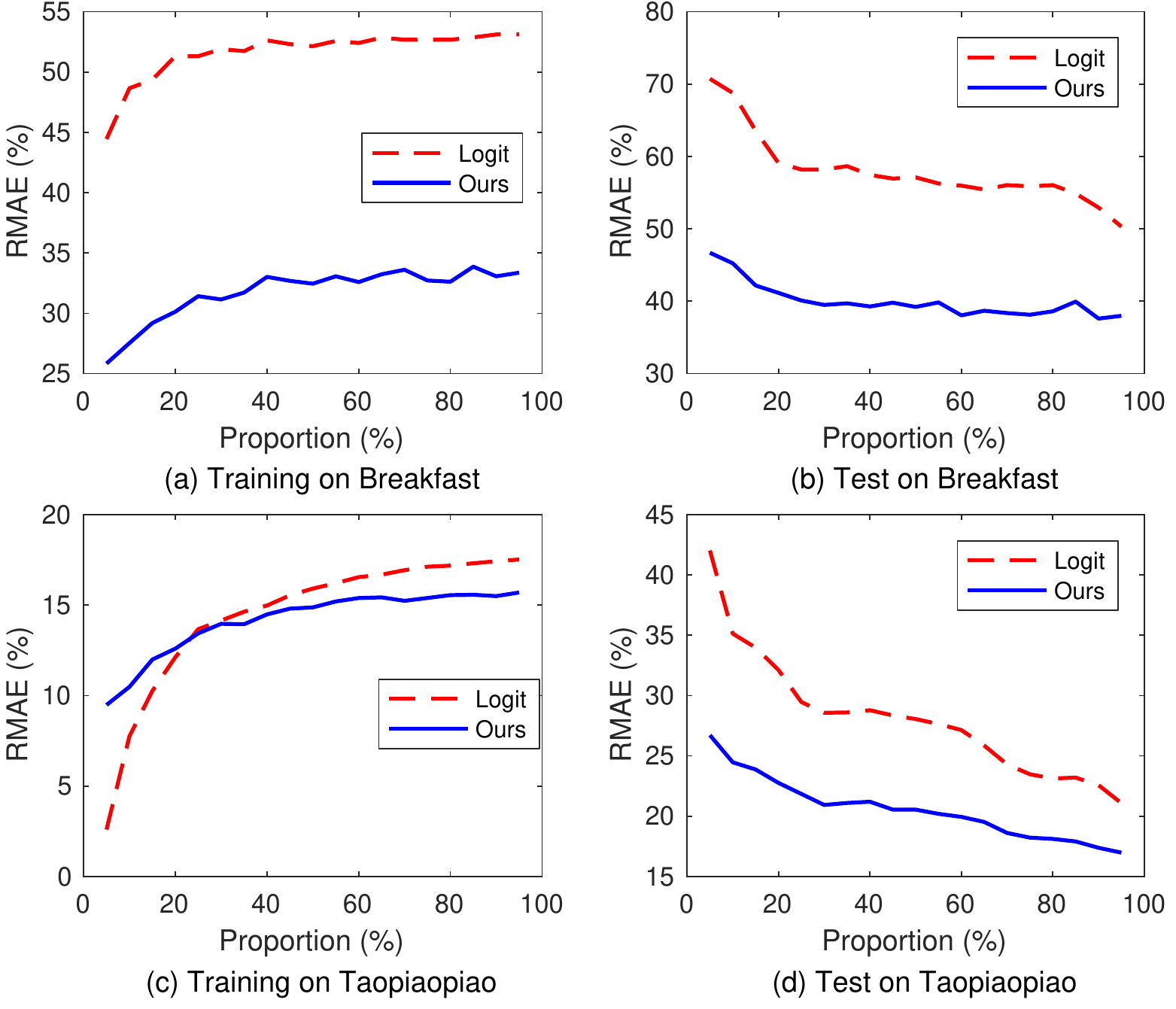}
	\caption{The comparisons between logit model 
	and our semi-black-box model under different partition ratios.}
	\label{fig:tpp_breakfast}
\end{figure}

\subsection{Budget Allocation}
To investigate the behaviors of our budget allocation algorithm, 
several simulations are conducted. 

\subsubsection{Setups}
We consider a synthetic scenario including $N=100$ market-segments, 
where the parameters of market response models are generated randomly. 
More specifically, the parameters are generated as follow: 
for $i=1,\dots, N$, $D_i \sim \mathcal{U}(0, 100), a_i \sim \mathcal{U}(-1, 1), b_i \sim \mathcal{U}(0, 1)$ 
and $B \sim \mathcal{U}(0, 100N)$, where $\mathcal{U}(l, u)$ represents the uniform distribution on interval $(l, u)$.

\subsubsection{Convergence}
The convergence processes of three random examples are shown in \figurename{ \ref{fig_converge}. 
We can see that the proposed algorithm converges in a few iterations, usually less than 10 iterations. 
As $g(\lambda)$ goes to $0$, the objective value approaches its maximum value, 
which confirms the derivation in Section \ref{sec:alg}. 
\begin{figure}[!t]
	\centering
	\includegraphics[width=0.93\linewidth]{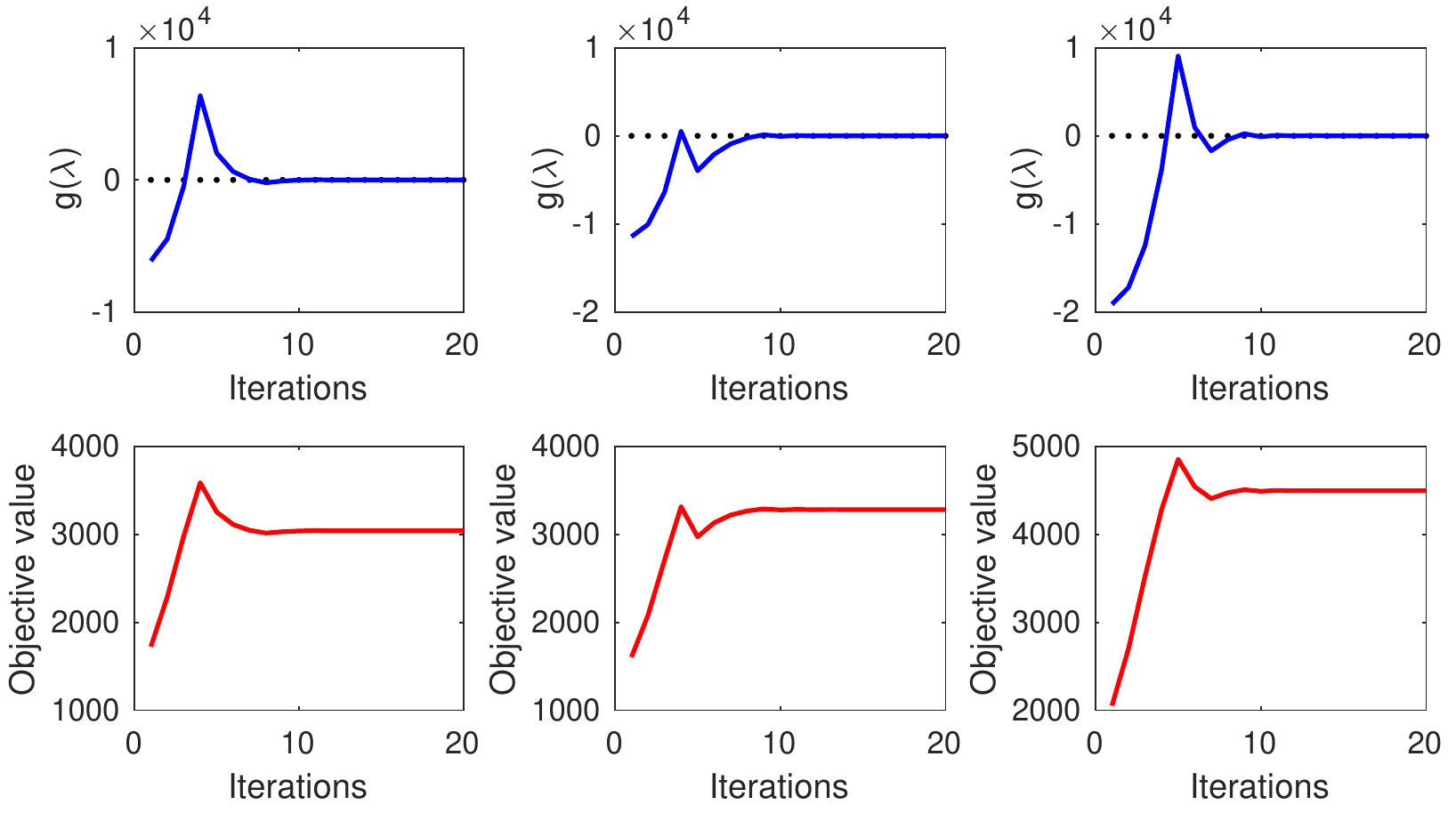}
	\caption{Convergence processes of three random examples (grouped by column).}
	\label{fig_converge}
\end{figure}

\subsubsection{Parameter sensitivity}
To study how the estimation bias of $a_i$ and $b_i$ disturbs our algorithm, 
some artificial biases are simulated on the generated parameters.

Firstly, by adding different level biases to the generated $a_i, b_i (i=1,\dots,N)$, 
we get the biased parameters: 
\begin{equation}
\hat{a}_i = a_i + \epsilon |a_i|, \ \hat{b}_i = b_i + \epsilon |b_i|,
\end{equation}
where $\epsilon$ is randomly sampled from $\mathcal{U}(-0.2, 0.2)$ for each group $a_i, b_i (i=1,\dots,N)$. 
Secondly, solution $\hat{c}_i$ is calculated under $(\hat{a}_i, b_i)$ and $(a_i, \hat{b}_i)$ respectively. 
Based on $\hat{c}_i, a_i, b_i$, objective value and constraint function (or exceeded budget) are 
calculated as: 
\begin{equation}
\hat{d}(\hat{\bm{c}}) = \sum_{i=1}^N D_i q_i(\hat{c}_i; a_i, b_i), \ \hat{g}(\hat{\bm{c}}) = \sum_{i=1}^N D_i q_i(\hat{c}_i; a_i, b_i)\hat{c}_i -B.
\end{equation}
Then we compute the relative error of objective value and exceeded budget as follow:
\begin{equation}
\begin{split}
&\mbox{Objective error} = \frac{d(\bm{c}^*) - \hat{d}(\hat{\bm{c}})}{d(\bm{c}^*)},  \\
&\mbox{Exceeded budget} = \frac{\hat{g}(\hat{\bm{c}}) }{B}.
\end{split}
\end{equation}
More than $1000$ independent Monte Carlo simulations are conducted, 
and the results are shown in \figurename{ \ref{fig_sensitive}. 
 
\begin{figure}[!t]
	\centering
	\includegraphics[width=0.93\linewidth]{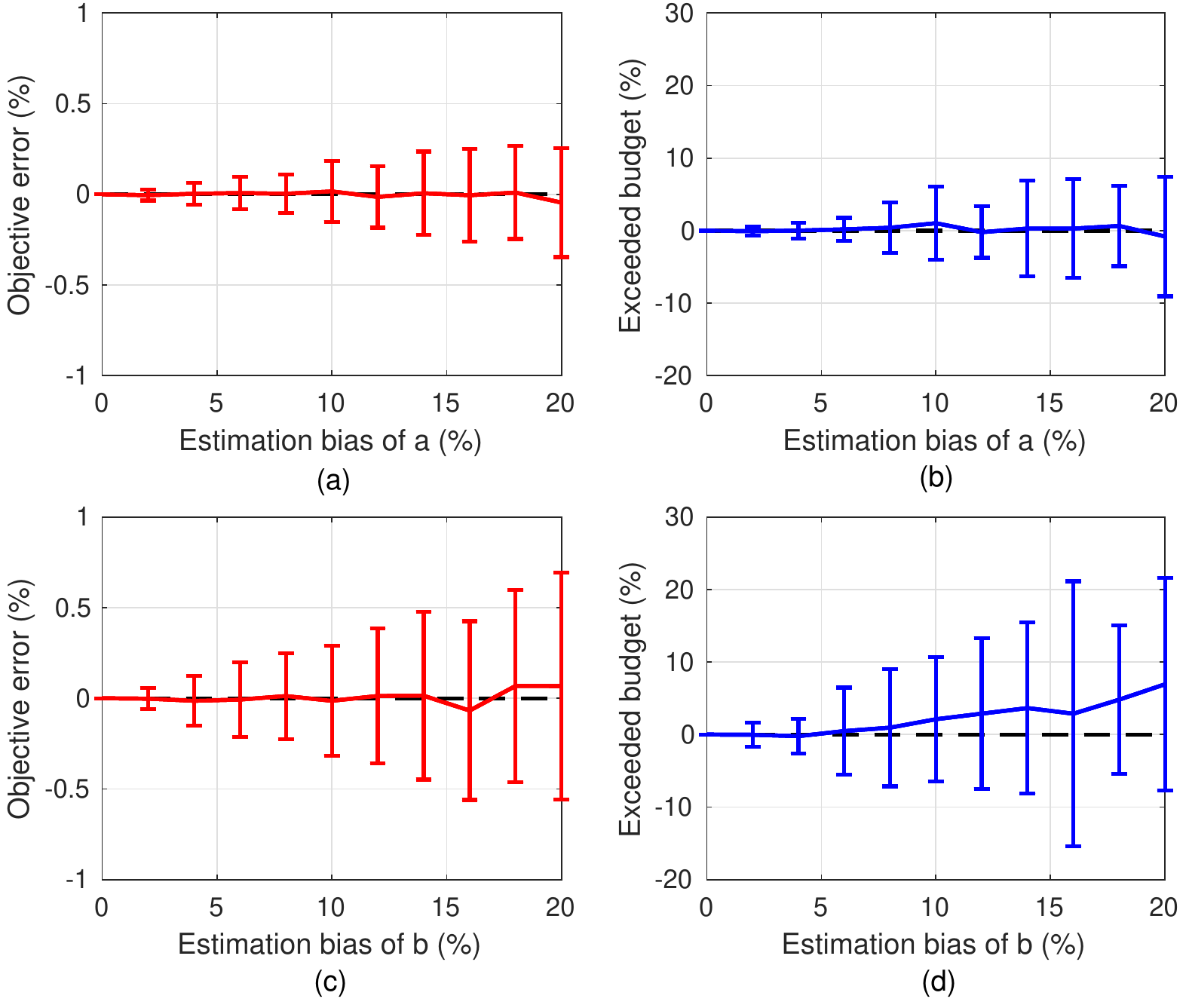}
	\caption{Parameter sensitivity.  (a) (b) The mean and standard deviation of objective error and 
	exceeded budget under different estimation biases of $a$. 
	(c) (d) The mean and standard deviation of objective error and exceeded budget
	under different estimation biases of $b$.}
	\label{fig_sensitive}
\end{figure}

Obviously, when the estimation bias increases, both objective error and exceeded budget increase.
However, the objective is not very sensitive to parameters, 
e.g. the objective error is still less than $1\%$ even with $20\%$ estimation bias.
On the other hand, although budget is more sensitive to parameters than objective, 
the exceeded budget is less than $5\%$ when estimation biases of $a_i, b_i$ are less than
 $10\%, 5\%$ respectively. Moreover, the algorithm is less sensitive to $a_i$ than $b_i$. 
One reason is that, in logit response models, $b_i$ is the coefficient of decision variable $c$ 
while $a_i$ is the bias term. 
Note that estimation bias is limited under a reasonable forecasting accuracy. 

\subsubsection{Discrete settings} 
The discrete constraints are common in many real-world situations, 
and we have presented efficient approximation algorithms for solving discrete problems. 
To evaluate the approximate error of proposed methods, several simulations are conducted. 
Two approximation algorithms are tested in simulations: 
1) Bisection + Dyer-Zemel, which solves the relaxed continuous problem first 
and then solves the constructed MCKP by Dyer-Zemel; 
2) Dyer-Zemel only, which is suggested when the set of options is small. 
As the distance between two adjacent optional values, which indicates 
the density in discrete constraints, 
varies from $0.1$ to $8$, $100$ independent Monte Carlo simulations are run for each density. 

The upper bound of approximate error can be calculated as follow: 
\begin{equation}
\mbox{Approximate Error} = \frac{d_{u} - d_a}{d_{u} - d_{0}},
\end{equation}
where $d_a$ is the objective value of approximate solution, 
$d_0$ is the objective value with no action (i.e. $c_i=0$ for every $i$),  
and $d_u$ is the optimal objective value of relaxed continuous problem. 
The detailed simulation results are shown in Table \ref{tab_discrete}, 
which includes both mean and standard derivation of approximate errors. 
It is shown that the upper bound of approximate error is very small when the options 
have a high density (i.e. small discrete distance). 
For instance, when discrete distances are less than $1$, 
the approximate error is no more than $1\%$.
As optional values become sparse, the upper bound of approximate error increases, 
which is consistent with our intuition. 
However, the real approximate error, 
which is unknown due to NP-hardness, is not necessarily significant. 
Moreover, the budget allocation 
results of two algorithms are the same in most situations. Therefore, 
it is reasonable to suggest directly solving problems with discrete constraints 
by Dyer-Zemel algorithm when the set of options is small 
and the running time is not the bottleneck. 

\begin{table}[!t]
	\centering
	\caption{The upper bound on approximate error of two algorithms for solving discrete problems.} 
	\label{tab_discrete}
	\begin{tabular}{c|c|c}
		\hline
		& \multicolumn{2}{c}{Approximate error (mean $\pm$ stddev) (\%)}   \\ \hline
		Distance & Bisection + Dyer-Zemel & Dyer-Zemel only \\ \hline \hline
		$0.1$  &   $0.0297 \pm 0.0337 $ & $ 0.0299 \pm 0.0337$ \\ 
		$0.5$  & $0.3040 \pm 0.2614$ & $0.3043 \pm 0.2612$ \\ 
		$1$  &   $0.8418 \pm 0.5582$ & $0.8418 \pm 0.5582$ \\ 
		$2$  &  $3.053 \pm 2.247$ & $3.053 \pm 2.247$ \\ 
		$4$  &  $15.14 \pm 10.43$ & $15.09 \pm 10.45$ \\ 
		$8$ & $49.89 \pm 11.52$ & $49.88 \pm 11.53$ \\ \hline
	\end{tabular}
\end{table}

\subsection{Real-world Scenarios}

In order to show the performance of our approach in real-world scenarios, 
we would like to take the daily marketing in Taopiaopiao as a typical example. 
Before policy changes, there are several billion RMB spent on marketing each year 
to compete for market shares (as equal to sales) with other platforms, Maoyan for an example. 
How to spend money efficiently is a crucial part in the competition. 

\subsubsection{Setups}
In order to avoid price discrimination, the unit cost (price discount or premium) on marketing
is determined at each market-segment (i.e. each day-of-week of different cinemas). 
For instance, discounts of movie tickets on Sunday for a specific cinema 
are the same for all Taopiaopiao customers, but the discounts can be different across 
different cinemas and also can be different from Saturday. 
The cost upper bound is updated weekly. 
Based on initial budget, the allocation is firstly 
calculated at the beginning of the following week. 
Then the allocation need to be readjusted daily based on the left budget.  
One major reason for budget allocation shifting is the biased forecasting. 
In addition, the discrete constraints are required here due to customers' preference of 99-ending prices. 

We design the following cinema-based randomized A/B testing experiment 
to form fair comparisons. 
Firstly, we filter out outliers, including cinemas with extreme high or 
low market shares in history, or abnormal variance of sales, etc. 
Secondly, 4 hundred cinemas are randomly picked across different cities as group A,  
and then a one-to-one mapping approach is conducted to generate group B. 
In the one-to-one mapping approach, the 1-nearest-neighbor is selected for each cinema in group A 
based on historical data and contextual features. Thirdly, an A/A testing runs for 
several days to confirm group A and B have similar sales volumes and marketing responses. 
Finally, an A/B testing runs for five weeks, where cost upper bounds of group A and B 
for each week are the same. The budget is uniformly allocated in group A, 
and allocated based on our approach in group B.

\subsubsection{Results}
The comparison results on market share are shown in \figurename{ \ref{img_performance_ratio}}, 
and the comparison results on ROI are shown in \figurename{ \ref{img_performance}}. 
First of all, the sales is effectively improved every week and the average improvement is above 6\%. 
Actually, the algorithm is always early terminated by $f(\lambda_l) - f(\lambda_r) \le \epsilon'$. 
The reason is that the cost upper bound is given 
by managers according to their subjective experience, 
and it is not necessary to spend that much money for negligible improvements. 
So our approach decides to spend much less money, and thus the improvements on ROI 
is remarkable (more than 40\% averagely). In one word, the proposed approach 
can improve sales over 6\% with spending 40\% less money on marketing 
by allocating budget into the right market-segments. 
As this example scenario shows, the data-driven methods can help us to make more informed 
and effective decisions in online business, where it is difficult for human to monitor and 
analyze the dynamic business environment constantly.  

\begin{figure}[t]
  \centering
  \includegraphics[width=0.93\linewidth]{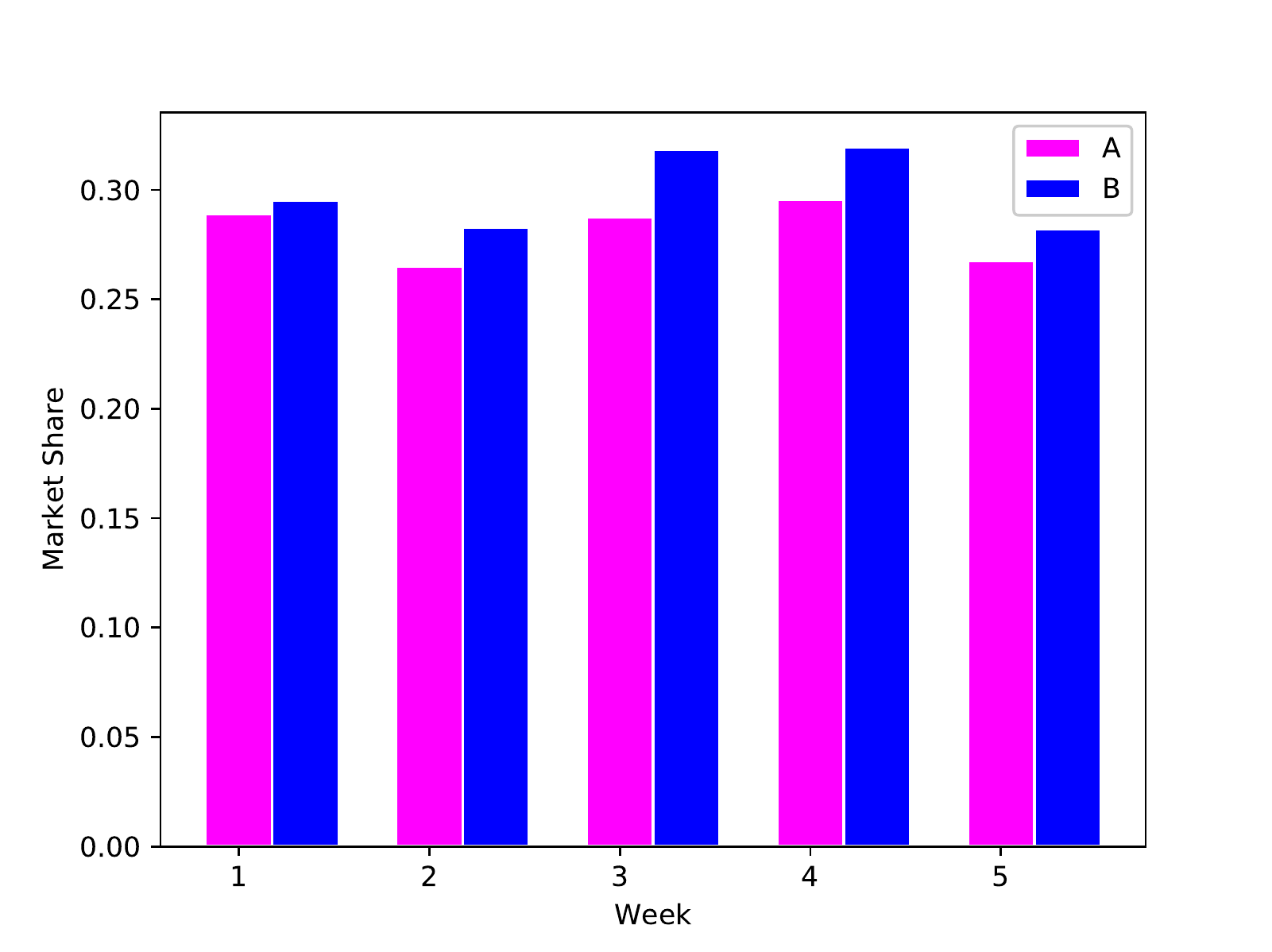} 
  \caption{Comparison results on market share.} 
  \label{img_performance_ratio} 
  \end{figure}

 \begin{figure}[t]
  \centering
  \includegraphics[width=0.93\linewidth]{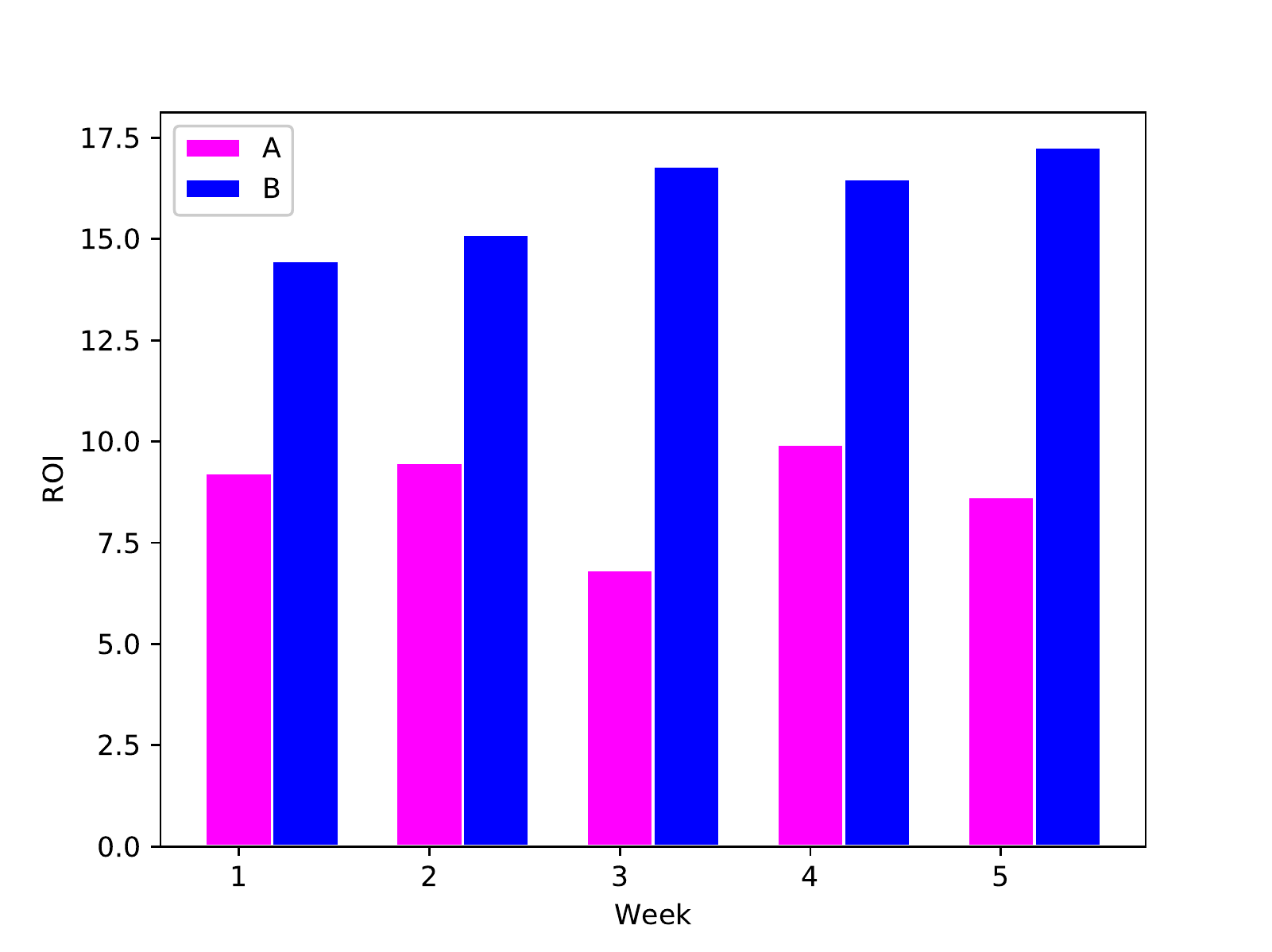} 
  \caption{Comparison results on ROI.} 
  \label{img_performance} 
  \end{figure}

%% file: related_work.tex
\section{Related Work}
Marketing is one of the premier components of business management, 
and it has been studied for decades \cite{malhotra2007review}. 
In marketing, companies need to allocate whole budget across different market-segments, 
such as across countries, products, consumption months and so on \cite{fischer2009dynamic}. 
There are great real benefits for improving the efficiency of marketing budget allocation, 
and thus many studies focus on this topic \cite{doyle1990multiproduct, reinartz2005balancing, 
gupta2008allocating}. Recently, online business has become an important part of our daily lives.  
The environment in online business is very dynamic and complex, 
which brings great challenges to marketing, especially for companies 
with large and diversified markets \cite{laudon2016commerce, strauss2016marketing}. 
Many data-driven techniques are developed \cite{akter2016big, ferreira2015analytics} to 
overcome these challenges, including forecasting \cite{beheshti2015survey, ye2018customized} 
and optimization \cite{ito2017optimization}. 

As black-box models, neural networks have recently been successfully applied 
to many forecasting applications \cite{tong2017simpler, liao2018deep, taieb2017regularization}. 
Through multiple non-linear transformations, they can achieve a high forecasting accuracy. 
However, there are several gaps between forecasting and decision making \cite{athey2017beyond}. 
One challenge is that black-box models often fail to identify relationship between 
the independent variable and objective variable, 
e.g. how cost on marketing effects sales, due to covariate shifting. 
A related topic on this problem is causal inference \cite{pearl2009causal, morgan2014counterfactuals}. 
Another challenge is that optimization on black-box functions is difficult. 
Therefore, economists prefer using explicit expressions to reveal relationship 
between the independent variable and objective variable. 
In the context of sales forecasting, the explicit models are called demand curves. 
There are many kinds of demand curves, such as linear, log-linear, 
constant-elasticity, logit and so on \cite{hanssens2003market, phillips2005pricing, talluri2006theory}. 
We focus on logit demand curve, which is the most popular response 
model \cite{phillips2005pricing, qu2013learning, van2018dynamic}. 
However, the capability of fully utilizing data is limited when the logit demand curve 
is fitted for each market-segment separately. That may cause serious 
data sparseness problem. One way to relieve data sparseness problem is 
Thompson sampling \cite{ganti2018thompson, ferreira2018online}, 
where independent variables need to change several time for explorations. 
In this work, we extend the capability of logit demand curves via neural networks 
and a semi-black-box model has been proposed to relieve the data sparseness problem. 

There are also massive works focusing on solving the marketing (or pricing) problems 
by formulating them as optimization problems. 
Ferreira et al. optimize pricing decisions by integer programming,
and solve it by linear relaxations \cite{ferreira2015analytics}. 
Ito et al. maximize future revenue or profit by binary quadratic programming,
and solve it by semi-definite relaxations \cite{ito2017optimization}. 
Boutilier et al. formulate the budget allocation as a 
sequential decision problem and solve it by MDP \cite{boutilier2016budget}. 
Staib et al. study the general budget allocation problem 
from a robust optimization perspective \cite{staib2017robust}.
In this work, the marketing budget allocation is firstly formulated as a non-convex problem. 
Inspired by \cite{dong2009dynamic, li2011pricing}, 
whose topics are pricing products with no budget constraints, 
we reformulate the problem into an equivalent convex optimization problem. 
Efficient algorithms are designed to solve it in both continuous and discrete settings. 
It is worth mentioning that a major cause of discrete decision variables is the 
preference of 99-ending display prices in many situations \cite{schindler1996increased}. 
The discrete settings can be solved by 
constructing a Multiple Choice Knapsack Problem (MCKP) 
from the solution of relaxed continuous problem. 
Dyer et al. \cite{dyer1984n} and Zemel et al. \cite{zemel1984n} independently 
developed approximation algorithms for MCKP, 
and the algorithm can be further boosted by pruning \cite{pisinger1995minimal}. 
Actually, MCKP has wide applications \cite{kellerer2004multiple}, 
such as advertising optimization in online marketing \cite{pani2017large}.

%% file: conclusion.tex
\section{Conclusion}
In this paper, we present a novel unified framework for marketing 
budget allocation, and there are two sequential steps in the proposed framework. 
Firstly, the response in each market-segment is forecasted by 
a semi-black-box model, which extends logit demand curves 
with the capability of neural networks. The new response model 
not only can share information across different segments to relieve 
data sparseness problem, but also keeps the explicit expression 
between the independent variable and objective variable. 
Then budget allocation is formulated as an optimization problem 
based on learned response models. The optimization problem can 
be solved efficiently by binary search on a dual variable in a few iterations 
and the complexity of each iteration is $O(N)$. 
Thus it is readily to handle large-scale problems. 
Several business constrains, including cost upper bound, profit lower bound, ROI lower bound, 
and discrete constrains, are supported in one unified paradigm with minor adaptations. 
This framework is easy to implement and has been successfully applied to many 
scenarios in Alibaba Group. The results of both offline experiments and 
online A/B testing demonstrate its effectiveness.  
Taking the daily marketing in Taopiaopiao as an example, 
the proposed approach can improve sales over 6\% 
with spending 40\% less money on marketing.

For future works, we are interested in exploring relationship 
between the market cost and contextual variables in the logit response model. 
There is also a strong motivation to support boundary constraints 
on decision variables in the optimization part. Moreover, 
another interesting topic is to theoretically analyze 
the approximation error of algorithms for solving the discrete setting.

%% file: appendix.tex
\section{Supplement}
\subsection{Implementation} 
In order to improve the reproducibility, we present more details about the implementation. 
\subsubsection{Software versions}
The details of languages, packages and frameworks used in the deployment and experiments are as follow:
\begin{itemize}
\item Language: Python 2.7.15, Cython 0.28.5, SQL. 
\item Packages: NumPy 1.15.4, SciPy 1.1.0, Tensorflow 1.11.0. 
\item Frameworks: MaxCompute, Spark 2.3.0.
\end{itemize}

\subsubsection{Large scale}
When the training set of forecasting is extremely large, batch gradient descent should be used 
instead of gradient descent, and machines with GPUs are preferred. 
However, the inference part can be easily paralleled. 
For the optimization part, our algorithm can be naturally implemented in a 
distributed framework, such as Spark, where computations are paralleled across market-segments. 

\subsubsection{Other tips}
\begin{itemize}
\item The experimental results on semi-black-box models are 
quite insensitivity to hyper-parameters in neural networks, 
and more layers or higher dimensions are not necessary. 
\item In some extreme cases, it may need to use high-precision computing to 
calculate $\bm{q}$ in Eq. (\ref{new_kkt_dual}) to avoid overflow caused by the exponent. 
The extreme cases only happen when $\epsilon$ is set to be a very small value, 
but it is not necessary in most situations. 
\item The pseudocode in page 323 of \cite{kellerer2004multiple} is a good reference 
for implementing the Dyer-Zemel algorithm. However, there are two misprints: 
the condition in step 6 should be $\sum_{i=1}^{m}w_{ia_i} > c$ 
and the condition in step 7 should be $\sum_{i=1}^{m}w_{ia_i} \le c$. 
\end{itemize}

\subsection{Proofs}
\subsubsection{Theorem \ref{theor_elasticity}}
\begin{proof}
The derivative of Eq. (\ref{eq:logit}) is 
\begin{equation}
\label{eq:logit_d}
\nabla_c d_i(c) = \frac{D_i b_i \exp{\{-(a_i + b_i c)\}}}{[1+\exp{\{-(a_i + b_i c)\}}]^2}. 
\end{equation}

Let us substitute it into Eq. (\ref{elasticity2}) and we can get:
\begin{equation}
e_i(c)=\frac{b_i c}{1+\exp{\{a_i + b_i c\}}}, 
\label{eq:elasticity_logit}
\end{equation}
then we have $e_i(\hat{c}_i)=e_i(-\frac{a_i}{b_i})=-\frac{a_i}{2}$. 
\end{proof}

\subsubsection{Theorem \ref{theor_q_convex}}
\begin{proof}
Note that 
\begin{equation}\label{dg_dq}
\frac{d g}{d q_i} = \frac{D_i}{b_i} (1-a_i + \ln \frac{q_i}{1-q_i} + \frac{q_i}{1-q_i}), i=1, ..., N,
\end{equation}
and then 
\begin{equation}
\frac{d^2 g}{d^2 q_i} = \frac{D_i}{b_iq_i(1-q_i)^2}, i=1, ..., N.
\end{equation}
That is $\nabla_{\bm{q}}^2 g(\bm{q}) = \text{diag}([\frac{D_i}{b_iq_i(1-q_i)^2}]) $. 
Since $D_i >0, b_i >0$ and $ 0<q_i < 1$,
we have $ \nabla_{\bm{q}}^2 g(\bm{q})  \succeq \text{diag}([\frac{27D_i}{4b_i}]) \succ 0$, 
where equality holds if and only if $q_i=\frac{1}{3}, \forall i$. 
Thus, $g(\bm{q})$ is \emph{strongly convex}.
\end{proof}

\subsubsection{Theorem \ref{theor_kkt_equvi}}
\begin{proof}
	From Eq. (\ref{kkt_primal}) and $D_i > 0$, we can get: 
	\begin{equation} \label{derivate_q}
		\lambda^*  (1- a_i +  \ln \frac{q_i^*}{1-q_i^*} +  \frac{q_i^*}{1-q_i^*}) = b_i.
	\end{equation}
	Since $b_i > 0$, we have $\lambda^* \neq 0$. 
	
	Let us define $x_i  := \frac{q_i^*}{1-q_i^*}$ and substitute it into (\ref{derivate_q}), we have
	\begin{equation}
		\ln x_i  +  x_i = a_i + \frac{b_i}{\lambda^*} - 1,
	\end{equation}
	or equivalently, 
	\begin{equation}
	x_i e^{x_i}= \exp (a_i + \frac{b_i}{\lambda^*} - 1).
	\end{equation}
	Since $\exp (a_i + \frac{b_i}{\lambda^*} - 1) > 0$, we have 
	\begin{equation}
	x_i = W \big(\exp (a_i + \frac{b_i }{\lambda^*} - 1) \big),
	\end{equation}
	where $W(\cdot)$ is the Lambert $W$ function \cite{corless1996lambertw}. 
	Then we can obtain $q_i^*  = \frac{x_i }{x_i +1} = \frac{W \big(\exp (a_i  + \frac{b_i }{\lambda^* } - 1) \big)}{W \big(\exp (a_i  + \frac{b_i }{\lambda^* } - 1) \big)+1}$. 
\end{proof}

\subsubsection{Theorem \ref{theor_fg_decreasing}}
\begin{proof}
	Firstly, let $x_i :=\frac{q_i}{1-q_i}$. According to 
	\begin{equation} 
	\begin{aligned}
	\frac{d \lambda}{d q_i}  & = \frac{d \lambda}{d x_i}  \frac{d x_i}{d q_i} \\
	& =\frac{-b_i (\frac{1}{x_i} + 1)}{(1 - a_i + \ln x_i + x_i)^2}  \frac{1}{(1-q_i)^2}\\
	& =  -\frac{\lambda^2}{b_i q_i (1-q_i)^2}, i=1,\dots,N,
	\end{aligned}
	\end{equation}
	we have: 
	\begin{equation}\label{derivate_q_lambda}
	\frac{d q_i}{d \lambda} = -\frac{b_i q_i (1-q_i)^2}{\lambda^2},i=1,\dots,N.
	\end{equation}
	The derivative of $f(\lambda)$ with respect to $\lambda$ is:
	\begin{equation}
	\begin{aligned}
	\frac{d f}{d \lambda} & = \sum_{i=1}^{N} D_i \frac{d q_i}{d \lambda} \\
	& = -\sum_{i=1}^{N} \frac{D_i b_i q_i (1-q_i)^2}{\lambda^2} < 0,
	\end{aligned}
	\end{equation}
	where $D_i > 0, b_i>0$ and $0<q_i <1$. Thus $f(\lambda)$ 
	is strictly decreasing with respect to $\lambda$.
	
	Secondly, based on Eq. (\ref{kkt_primal}) we have: 
	\begin{equation}
	1- a_i +  \ln x_i + x_i = \frac{b_i}{\lambda}
	\end{equation}
	Then according to Eq. (\ref{dg_dq}), we have:
	\begin{equation}
	\frac{d g_i}{d  q_i} = \frac{D_i}{\lambda}, i=1, ..., N.
	\end{equation}
	The derivative of $g(\lambda)$ with respect to $\lambda$ is:  
	\begin{equation}
	\begin{aligned}
	\frac{d g}{d \lambda} & = \sum_{i=1}^{N} \frac{d g_i}{d \lambda} = \sum_{i=1}^{N} \frac{d g_i}{d q_i} \frac{d q_i}{d \lambda} \\
	& = -\sum_{i=1}^{N}  \frac{D_ib_i q_i (1-q_i)^2}{\lambda^3} < 0, 
	\end{aligned}
	\end{equation}
	where $D_i > 0, b_i>0$ and $0<q_i <1$. Thus $g(\lambda)$ 
	is strictly decreasing with respect to $\lambda$.
	
	Finally, since $\lim\limits_{\lambda \to 0} q_i(\lambda) = 1$ and $\lim\limits_{q_i \to 1} c_i(q_i) = +\infty, i=1,\dots,N$, 
	we have $\lim\limits_{\lambda \to 0} g(\lambda) \to +\infty > 0$.
\end{proof}

\subsubsection{Theorem \ref{theor_ext_convex}}
\begin{proof}
Firstly, $\nabla_{\bm{q}}^2 g'(\bm{q}) = \text{diag}([\frac{R D_i}{b_iq_i(1-q_i)^2}]) $. 
Since $R > 0, D_i >0, b_i >0$ and $ 0<q_i < 1$,
we have $ \nabla_{\bm{q}}^2 g'(\bm{q})  \succeq \text{diag}([\frac{27R D_i}{4b_i}]) \succ 0$, 
where equality holds if and only if $q_i=\frac{1}{3}, \forall i$. 
Thus, $g' (\bm{q})$ is \emph{strongly convex}.

Secondly, the derivative of $g'(\lambda)$ with respect to $\lambda$ is:  
	\begin{equation}
	\frac{d g'}{d \lambda} = -\sum_{i=1}^{N}  \frac{D_i b_i q_i (1-q_i)^2}{R \lambda^3} < 0, 
	\end{equation}
	where $R > 0, D_i > 0, b_i>0$ and $0<q_i <1$. Thus $g'(\lambda)$ 
	is strictly decreasing with respect to $\lambda$.
	
Finally, since $\lim\limits_{\lambda \to 0} q_i(\lambda) = 1$ and $\lim\limits_{q_i \to 1} c_i(q_i) = +\infty, i=1,\dots,N$, 
	we have $\lim\limits_{\lambda \to 0} g'(\lambda) \to +\infty > 0$.
\end{proof}